\documentclass[fleqn,usenatbib]{mnras}

\usepackage{newtxtext,newtxmath}

\usepackage[T1]{fontenc}
\usepackage{ae,aecompl}

\usepackage{ulem}

\usepackage{graphicx}	
\usepackage{amsmath}	
\usepackage{amssymb}	
\usepackage{subfig}
\newcommand*{\rom}[1]{\expandafter\@slowromancap\romannumeral #1@}
\newcommand*{\Rom}[1]{\expandafter\@slowromancap\romannumeral #1@}
\makeatother

\title[Short title, max. 45 characters]{Optical and X-ray study of the peculiar high mass X-ray binary XMMU J010331.7-730144}

\author[Monageng et al.]
  {I.M. Monageng$^{1,4}$\thanks{E-mail: itu@saao.ac.za}, 
  M.J. Coe$^{2}$,
  D.A.H. Buckley$^1$,
  V.A. McBride$^{1,4}$,
  J.A. Kennea$^3$, 
  \newauthor
  A. Udalski$^5$, 
  P.A. Evans$^6$,
  J.S. Clark$^8$,
  I. Negueruela$^7$
\\
$^1$South African Astronomical Observatory, P.O Box 9, Observatory, 7935, Cape Town, South Africa\\
$^2$Physics \& Astronomy, University of Southampton, SO17 1BJ, UK\\
$^3$Department of Astronomy and Astrophysics, The Pennsylvania State University, University Park, PA 16802, USA\\
$^4$Department of Astronomy, University of Cape Town, Private Bag X3, Rondebosch 7701, South Africa\\
$^5$Astronomical Observatory, University
of Warsaw, Al. Ujazdowskie 4, 00-478 Warszawa, Poland\\
$^6$University of Leicester, X-ray and Observational Astronomy Research Group, School of Physics \& Astronomy, University Road, Leicester LE1 7RH, UK\\
$^7$Dpto de Fisica Aplicada. Universidad de Alicante, Carretera de San Vicente del Raspeig s/n, 03690, Spain\\
$^8$Department of Physics and Astronomy, The Open University, Walton Hall, Milton Keynes, MK7 6AA, UK\\
}

\date{Accepted XXX. Received YYY; in original form ZZZ}

\pubyear{2019}

\begin{document}
\label{firstpage}
\pagerange{\pageref{firstpage}--\pageref{lastpage}}
\maketitle

\begin{abstract}
For a long time XMMU J010331.7-730144 was proposed as a high-mass X-ray binary candidate based on its X-ray properties, however, its optical behaviour was unclear - in particular previous observations did not reveal key Balmer emission lines. In this paper we report on optical and X-ray variability of the system. XMMU J010331.7-730144 has been monitored with the Optical Gravitational Lensing Experiment (OGLE) in the $I$ and $V-$bands for the past 9 years where it has shown extremely large amplitude outbursts separated by long periods of low-level flux. During its most recent optical outburst we obtained spectra with the Southern Africa Large Telescope (SALT) where, for the first time, the H$\alpha$ line is seen in emission, confirming the Be nature of the optical companion. The OGLE colour-magnitude diagrams also exhibit a distinct loop which is explained by changes in mass-loss from the Be star and mass outflow in its disc. In the X-rays, XMMU J010331.7-730144 has been monitored by the Neil Gehrels \textit{Swift} Observatory through the S-CUBED programme. The X-ray flux throughout the monitoring campaign shows relatively low values for a typical Be/X-ray binary system. We show, from the analysis of the optical data, that the variability is due to the Be disc density and opacity changing rather than its physical extent as a result of efficient truncation by the NS. The relatively low X-ray flux can then be explained by the neutron star normally accreting matter at a low rate due to the small radial extent of the Be disc. \\

\end{abstract}




\section{Introduction}

Be stars are early-type, non-supergiant stars of B spectral class that display, or have displayed Balmer emission lines, at some point in time, in their optical spectra \citep{1987pbes.coll....3C}. The emission lines originate from the matter which makes up the circumstellar disc around the B star. The formation of the disc has been a subject of scrutiny over the years and is thought to be due to a combination of the rapid rotation of the star and non-radial pulsations (e.g. \citealt{2003PASP..115.1153P,2009ApJ...701..396C}). \\
When a compact object is in orbit around a Be star, the system is referred to as a Be X-ray binary. Be X-ray binaries (BeXBs) make up the largest subclass of high mass X-ray binaries (HMXBs), with 49\% of the total population consisting of them \citep{2013ApJ...764..185C}. The interaction of the compact object, primarily a neutron star (NS), with the circumstellar disc results in accretion of matter leading to X-ray outbursts. The X-ray outbursts occur in two flavours: type I ($L \leq 10^{37}$erg $\cdot$ s$^{-1}$) and type II ($L \geq 10^{37}$erg $\cdot$ s$^{-1}$; \citealt{1986ApJ...308..669S}). For a general review of BeXBs, see \cite{2011Ap&SS.332....1R}.\\

In this paper we present analysis performed on XMMU J010331.7-730144 (X0103, hereafter). X0103 is in the \textit{XMM-Newton} Small Magellanic Cloud-survey point-source catalogue, where it was classified as a candidate HMXB \citep{2013A&A...558A...3S}. However, in a later study of the Small Magellanic Cloud (SMC), \cite{2016A&A...586A..81H} presented a complete catalogue of the dwarf galaxy where X0103 was deemed unlikely to be a HMXB, as no Balmer emission lines had been detected in the spectrum of the optical counterpart. We present the most comprehensive study of X0103 to date, where we use X-ray and optical data to analyse its historical and most recent behaviour. We use optical spectra to confirm the BeXB nature of X0103 by studying the Balmer line profiles. Furthermore, the long-term optical and X-ray data reveal unusual behaviour, with the optical photometric outbursts showing the largest amplitudes of all the known BeXB systems whilst the X-ray flux and H$\alpha$ equivalent width (EW) remain at a low level all the time. We explain this behaviour by arguing that the NS truncates the Be disc efficiently, thereby increasing its density and opacity. The truncation impedes the Be disc from growing to a large radial extent, resulting in a low accretion rate.\\
This paper is structured as follows: In section~\ref{sec:observations} we present the observations and analysis of the data. The results are presented in Section~\ref{sec:results}, while in Section~\ref{sec:discussion} we provide a discussion of the results. The conclusions are summarised in Section~\ref{sec:conclusion}.
\section{Observations}
\label{sec:observations}
\subsection{OGLE}
X0103 has been monitored with the Optical Gravitational Lensing Experiment (OGLE) project \citep{1997AcA....47..319U,2015AcA....65....1U} using the $I-$band filter over a period spanning $\sim$9 years with a cadence of $\sim$1-5 days. $V-$band observations have also been performed in the same time interval with a lower cadence ($\sim$1-400 days). The data reduction and calibrations were performed using the standard OGLE pipeline \citep{2015AcA....65....1U}. The lightcurves of the $I$ and $V-$band magnitudes are shown in Fig.~\ref{fig:ogle}.

\subsection{SALT}
The optical counterpart of X0103 was observed with the Southern African Large Telescope (SALT; \citealt{2006SPIE.6267E..0ZB}) using the Robert Stobie Spectrograph (RSS; \citealt{2003SPIE.4841.1463B,2003SPIE.4841.1634K}). The observations were performed using different gratings which cover different wavelength regions: PG0900 ($4350-7400$~\AA), PG1300 ($3800-6000$~\AA), PG1800 ($5985-7250$~\AA) and PG2300 ($6200-6800$~\AA). A summary of the observations is provided in Table \ref{tab:salt}. The SALT pipeline was used to perform the primary reductions which include overscan correlation, bias subtraction, gain correction and amplifier cross-talk corrections \citep{2012ascl.soft07010C}. The remaining reduction steps, which include wavelength calibration, background subtraction and extraction of the one-dimensional spectra were executed using \textsc{iraf}\footnote{Image Reduction and Analysis Facility: iraf.noao.edu}.

\begin{table}
	\centering
	\caption{A summary of the settings for the SALT observations}
	\label{tab:salt}
    \setlength\tabcolsep{0.8pt}
	\begin{tabular}{ccccc} 
		\hline\hline
		 & Grating  & Wavelength & Exposure &    \\ [-3pt]
		Grating &  angle &  range (\AA) & time (s) & Resolution (\AA)  \\
		\hline
		PG0900 & 15.125 & $4350-7400$  & 1200 & 6.10 \\
		PG1300 & 18.875 & $3800-6000$  & 900 & 4.00 \\
		PG1800 & 36.875 & $5985-7250$  & 1200 & 2.48 \\
		PG2300 & 48.872 & $6200-6800$  & 1000 & 1.68 \\
		\hline
	\end{tabular}
	\label{tab:SALT_regions}
\end{table}

\subsection{Swift}
\begin{figure*}
\resizebox{\hsize}{!}
            {\includegraphics[angle=0,width=17cm]{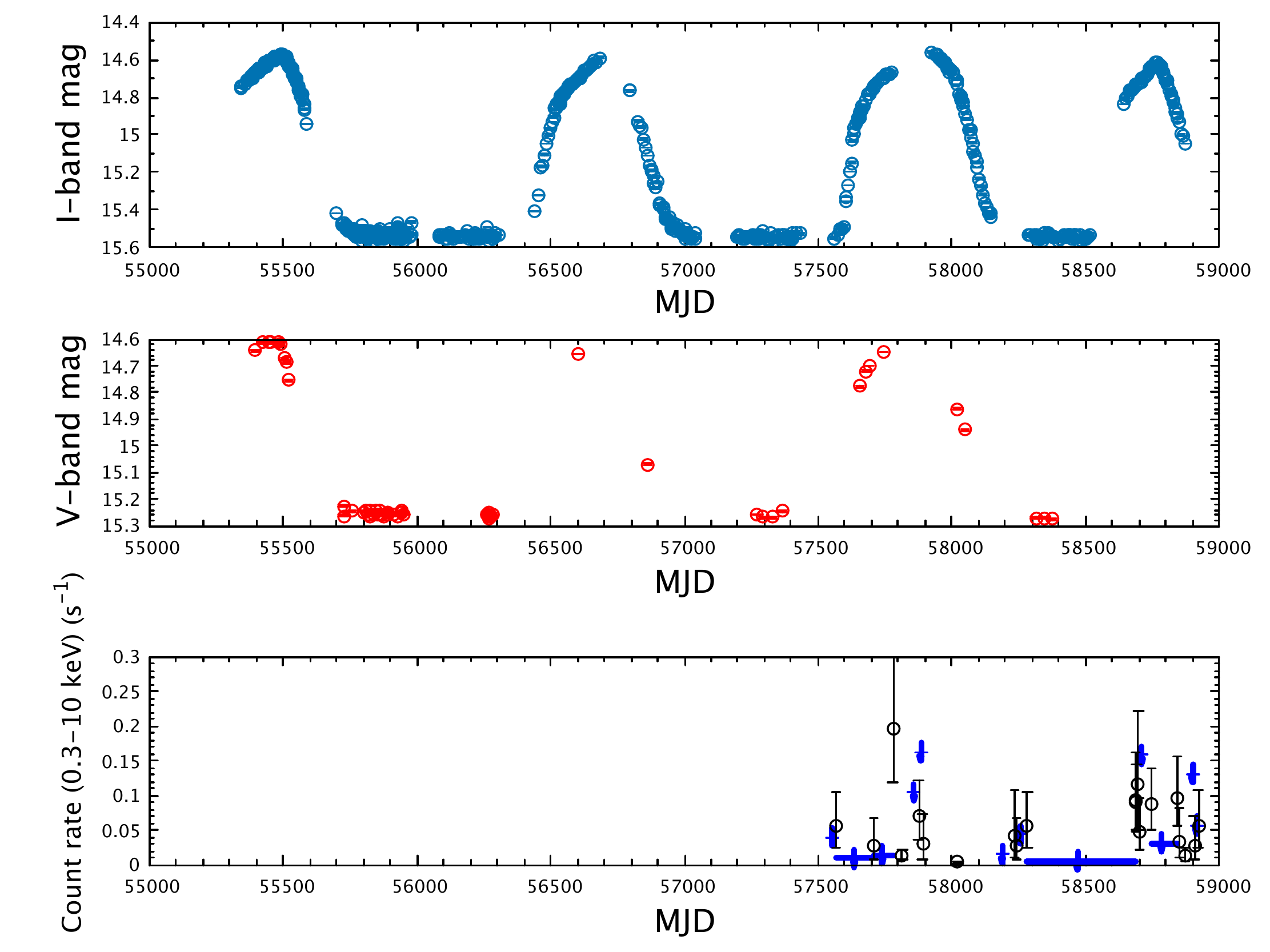}}
\caption{Long-term OGLE $I-$ (top panel) and $V-$band (middle panel) magnitude variability. The \textit{Swift} X-ray variability from the S-CUBED programme is shown in the bottom panel, where the blue arrows indicate the upper limits and the black circles show the detections. The red symbol is from the longer \textit{Swift} exposure.}
\label{fig:ogle}
\end{figure*}

X0103 has been observed $\sim$ weekly by the Neil Gehrels Swift Observatory (\textit{Swift}; \citealt{2004ApJ...611.1005G}) since 2016 June as part of the \textit{Swift} SMC Survey (S-CUBED) programme \citep{2018ApJ...868...47K}. The typical exposure times for these observations are approximately 60 seconds, with a total of 10.5 ks as of 2020 March 27. This results in a total of 122 observations. The data were analysed using an automated pipeline based on the tools of \cite{2009MNRAS.397.1177E}, as described in \cite{2018ApJ...868...47K}. The default light curve contains one bin per 60-s observation; however, in only 14 observations did this result in a significant (3-$\sigma$) detection. For the remaining 108 observations, we summed consecutive data points until either the combination resulted in a 3-$\sigma$ detection, in which case we created a light curve bin from these combined data; or until one of the 14 observations in which the source was detected was reached, in which case a 3-$\sigma$ upper limit was produced from the summed data prior to this point. 
The resultant light curve is shown in Fig.~\ref{fig:ogle} (bottom panel), where there are 20 flux measurements and 12 upper limits. The energy conversion factor used in this light curve was created from a spectrum formed from all observations to date. This can be modelled as an absorbed power-law with $N_H=7.1^{+84.2}_{-7.1} \times 10^{20}$ cm$^{-2}$, and photon index $\Gamma = 0.65^{+0.81}_{-0.48}$, yielding an energy conversion factor of $7.34 \times 10^{-11}$ erg cm$^{-2}$ ct$^{-1}$. 
The resulting $N_H$ is consistent with the column density to the SMC of $N_H=5.7 \times 10^{20}$ cm$^{-2}$ reported by \cite{1990ARA&A..28..215D}. We note that our spectral index is quite hard, but the uncertainties on that quoted value put it well within the range seen for other HMXBs. \cite{2005AstL...31..729F} in their review of spectral indices from X-ray accreting HMXBs found that the photon index, gamma, for such systems can lie anywhere in the range 0.3 - 2.3 depending upon the activity states of the systems. Our value falls well within the observed spread of values they quote. There was one deeper \textit{Swift} observation carried out on 13 Aug 2019 (MJD 58709) with an observation time of 1965s. The source was barely detected with a resulting count rate of $(8.1\pm2.0)  \times 10^{-3}$s$^{-1}$ - the lowest we recorded.


\begin{figure*}%
    \centering
    \subfloat[ ]{{\includegraphics[width=8cm]{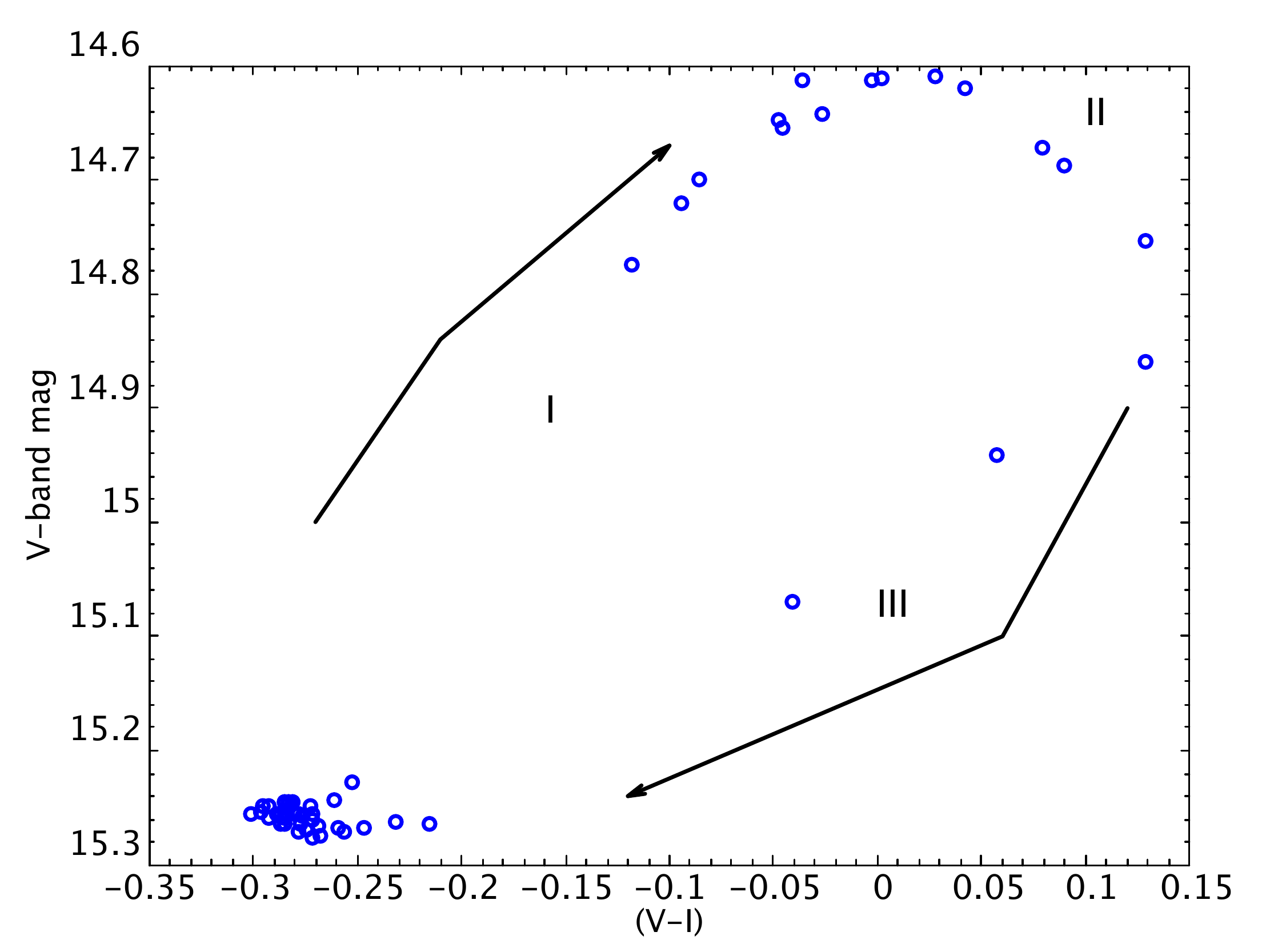} }}%
    \qquad
    \subfloat[ ]{{\includegraphics[width=8cm]{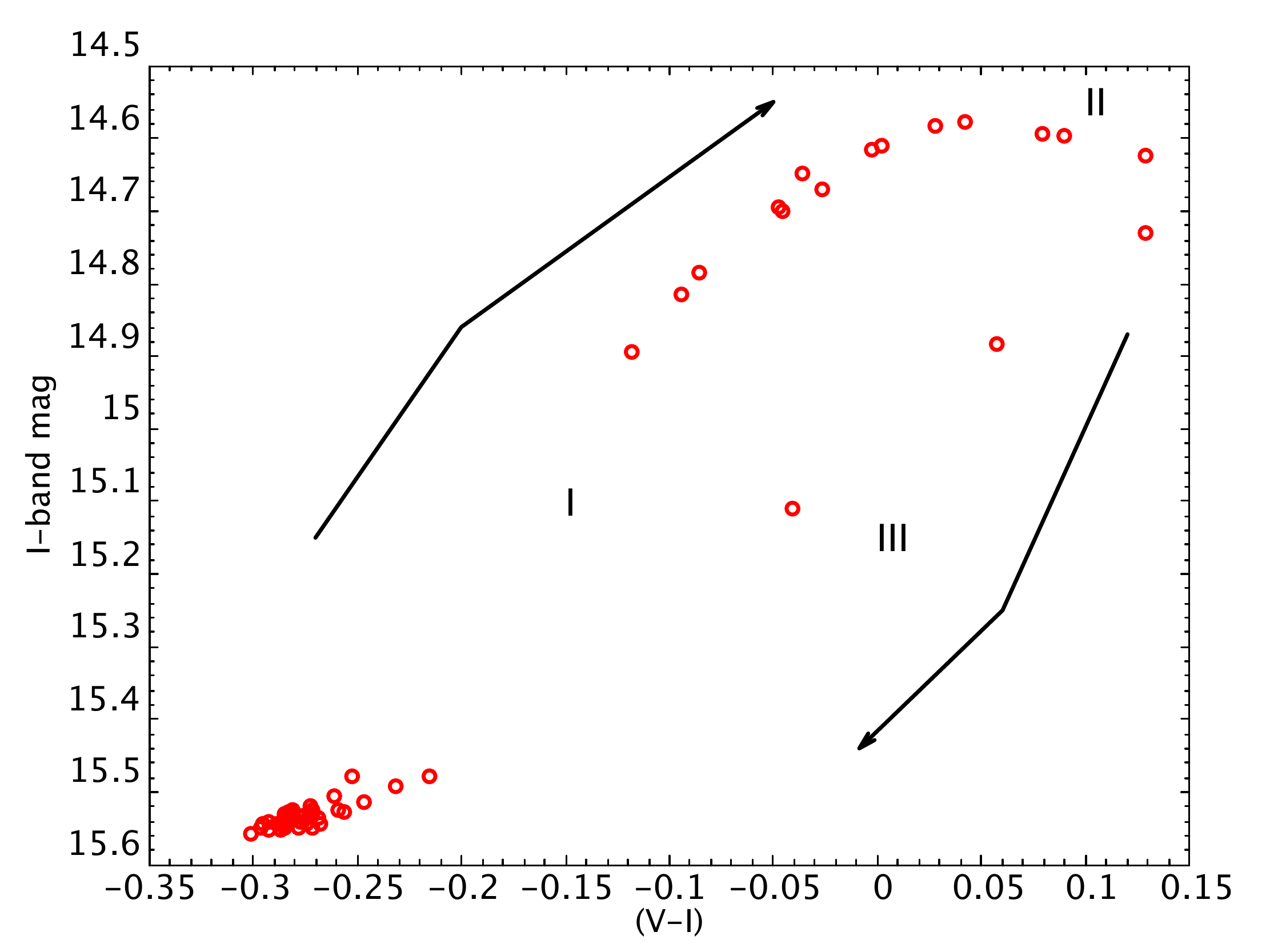} }}%
    \caption{$(V-I)-V$ (left) and $(V-I)-I$ (right) colour-magnitude plots from OGLE data. The arrows indicate the direction of evolution. Labels "I", "II" and "III" represent different phases in the disc evolution as explained in the text.}%
    \label{fig:col_mag}%
\end{figure*}

\section{Results}
\label{sec:results}
\subsection{Optical variability}
\label{sec:photometric_variability}
\subsubsection{Long term OGLE lightcurve}
Fig.~\ref{fig:ogle} shows the evolution of the OGLE $I$ and $V-$band magnitudes taken over a period of 9 years. The lightcurves reveal large outbursts with amplitudes of $\sim$1 and $\sim$0.6 magnitude in the $I$ and $V-$band, respectively, that occur regularly. To our knowledge, these are the strongest optical outbursts ever reported in a BeXB. The recurring timescale of the outbursts has a range $\sim 1150 - 1300$~days and last for $\sim 700-750$~days. This remarkable variability seen in the photometric lightcurves is believed to be an indicator of size and structural changes of the Be disc \citep{2011MNRAS.413.1600R}. While similar patterns in the photometric variability have been seen in BeXB systems such as CXO J005215.4$-$731915 \citep{2011MNRAS.412..391S,2019ApJ...884....2L}, the large amplitude of X0103 is atypical of the known systems.
\subsubsection{$(V-I)-I$ and $(V-I)-V$ colour-magnitude variability}


The $(V-I)-I$ and $(V-I)-V$ colour-magnitude variability is shown in Fig.~\ref{fig:col_mag}. The colour-magnitude plots shows a positive correlation with a loop structure. The positive correlation between the magnitude and colour is indicative of a low inclination angle of the disc relative to our line of sight \citep{1983HvaOB...7...55H,2011MNRAS.413.1600R,2015A&A...574A..33R}. This deduction from the relationship between the magnitude and colour follows from the fact that as the disc grows in size, the Be disc/star system gets brighter while the red continuum increases. Circumstellar discs in BeXBs are generally seen to appear redder as they grow in size since the outer disc regions are cooler in temperature than the star. For systems of high inclination angle (close to edge-on viewing), an anti-correlation would be seen in the evolution of the two quantities. In the edge-on scenario, as the disc grows in size the system reddens and gets dimmer, since the flaring disc obscures the light from the Be star. \\

The colour-magnitude plot shows a uniquely distinct looping structure which evolves in a clockwise direction. The loop structure can be explained by the interplay between mass-loss from the Be star and the outflow of the disc material \citep{2006A&A...456.1027D}. Phase "I" in Fig.~\ref{fig:col_mag} represents the stage when stellar mass-loss results in the production of the Be disc, which leads to an excess flux and reddening of the system until maximum flux is reached. Once the stellar mass outflow stops (when maximum flux is reached at the end of phase "I"), the magnitude begins to drop rapidly while the system is still reddening since the inner, bluer and optically thicker regions are removed first as the disc evolves to a ring (phase "II"). In phase "III" the colour moves towards the bluer values while the flux continues to decrease since the redder, optically thinner radiation is being removed faster. For a spectral type of B0 - O9.7~IV-V (see section~\ref{sec:spec_class}), a disc-less star would have an apparent $I-$band magnitude of $\sim15.2-15.4$ and a $(V-I)\sim -0.3$ colour index \citep{1981Ap&SS..80..353S,1993AcA....43..209W}. This places it towards the bottom left hand corner of Fig.~\ref{fig:col_mag}, suggesting that the disc around the optical counterpart in X0103 is almost completely depleted during minimum photometric emission at the end of the outbursts in Fig~\ref{fig:ogle} (top and middle panels).

\subsection{Optical spectroscopy}
\subsubsection{Spectral classification}
\label{sec:spec_class}
\begin{figure*}
\resizebox{\hsize}{!}
            {\includegraphics[angle=0,width=17cm]{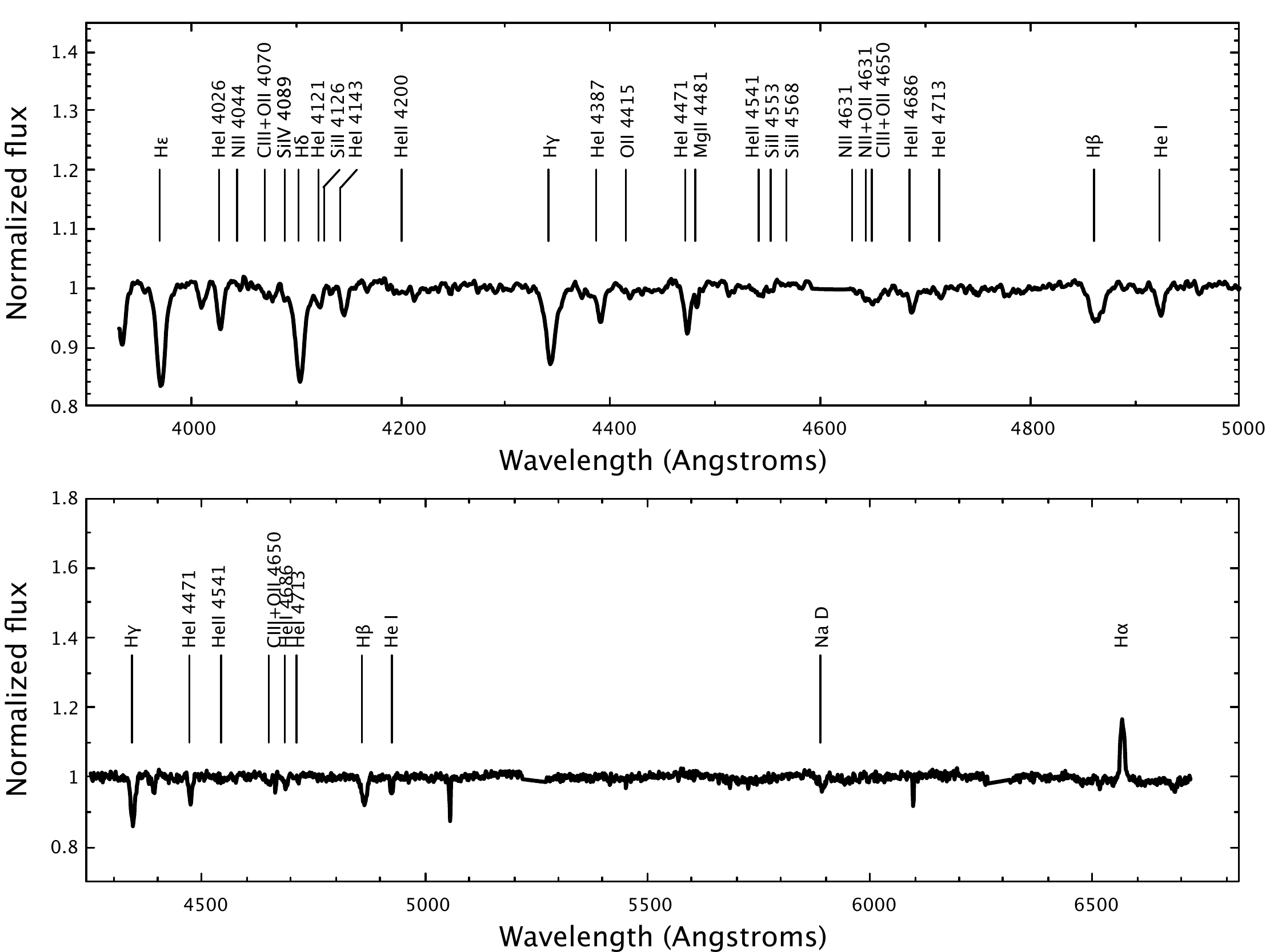}}
\caption{SALT spectra of X0103 taken with the PG1300 (top panel) and PG0900 gratings (bottom panel). The top spectrum covers the blue region and the bottom spectrum covers the full optical region. The different line species are labeled at their expected rest wavelengths.}
\label{fig:broad_band_spec}
\end{figure*}
Fig~\ref{fig:broad_band_spec} shows the blue (top panel) and broad-band (bottom panel) spectra obtained with SALT. The spectra are corrected for the heliocenter and the redshift of the SMC. The broad-band spectrum was taken on the 12th of August 2019 (MJD58708.02) with the PG0900 grating covering the blue and red regions of the optical waveband. As seen in the spectrum, there are a number of Balmer lines in absorption, indicative of an early spectral type. The spectrum also shows, for the first time, the H$\alpha$ line in emission. Previous observations covering the H$\alpha$ line have been obtained during optical minimum when the line was seen in absorption \citep{2016A&A...586A..81H}. The blue spectrum, taken with the PG1300 grating reveals the HeII 4686 absorption line, constraining the spectral type to B0 or earlier. The equal strength of the HeII 4541 and SiIII 4553 absorption lines further constrains the spectral type to O9.7. To obtain the luminosity class, we use the $V-$band magnitude range obtained from the OGLE data ($V\sim15.3-14.6$) with the distance modulus of the SMC, m$_V - $M$_V$ = 18.95 \citep{2013IAUS..289..222G}. This constrains the luminosity to a range IV-V. We conclude that the spectral type of the optical companion is O9.7 - B0 IV-V.  

\subsubsection{Be disc variability}
\begin{figure}
\resizebox{\hsize}{!}
            {\includegraphics[angle=0,width=17cm]{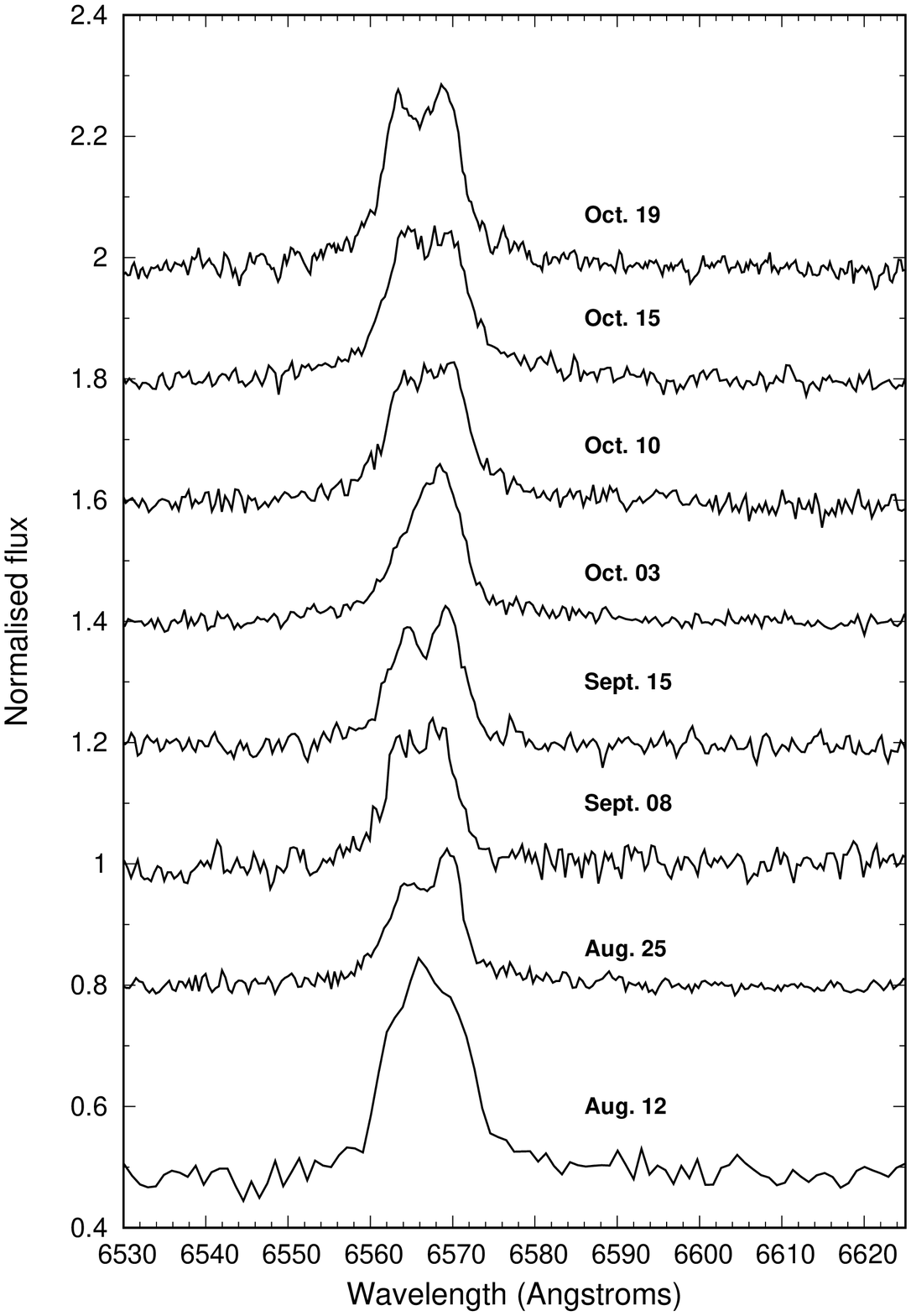}}
\caption{Evolution of the H$\alpha$ EW from SALT observations obtained in 2019 during the current OGLE outburst.}
\label{fig:H_alpha_evolution}
\end{figure}

We have obtained spectra covering the H$\alpha$ line region using the PG1800 and PG2300 gratings (see Table~\ref{tab:SALT_regions}). The spectra reveal the H$\alpha$ line present in emission since the commencement of the monitoring campaign. Fig.~\ref{fig:H_alpha_evolution} shows the evolution of the H$\alpha$ line morphology where the line profile is double-peaked due to the Keplerian distribution of disc matter viewed at non-zero inclination angles. The asymmetric single-peaked line profile obtained on the 12th of August 2019 is from an observation performed with the PG0900 grating, whose resolution is insufficient to resolve the two peaks. The EW measurements are logged in Table~\ref{tab:EW_measurements}.

\begin{figure}
\resizebox{\hsize}{!}
            {\includegraphics[angle=0,width=20cm]{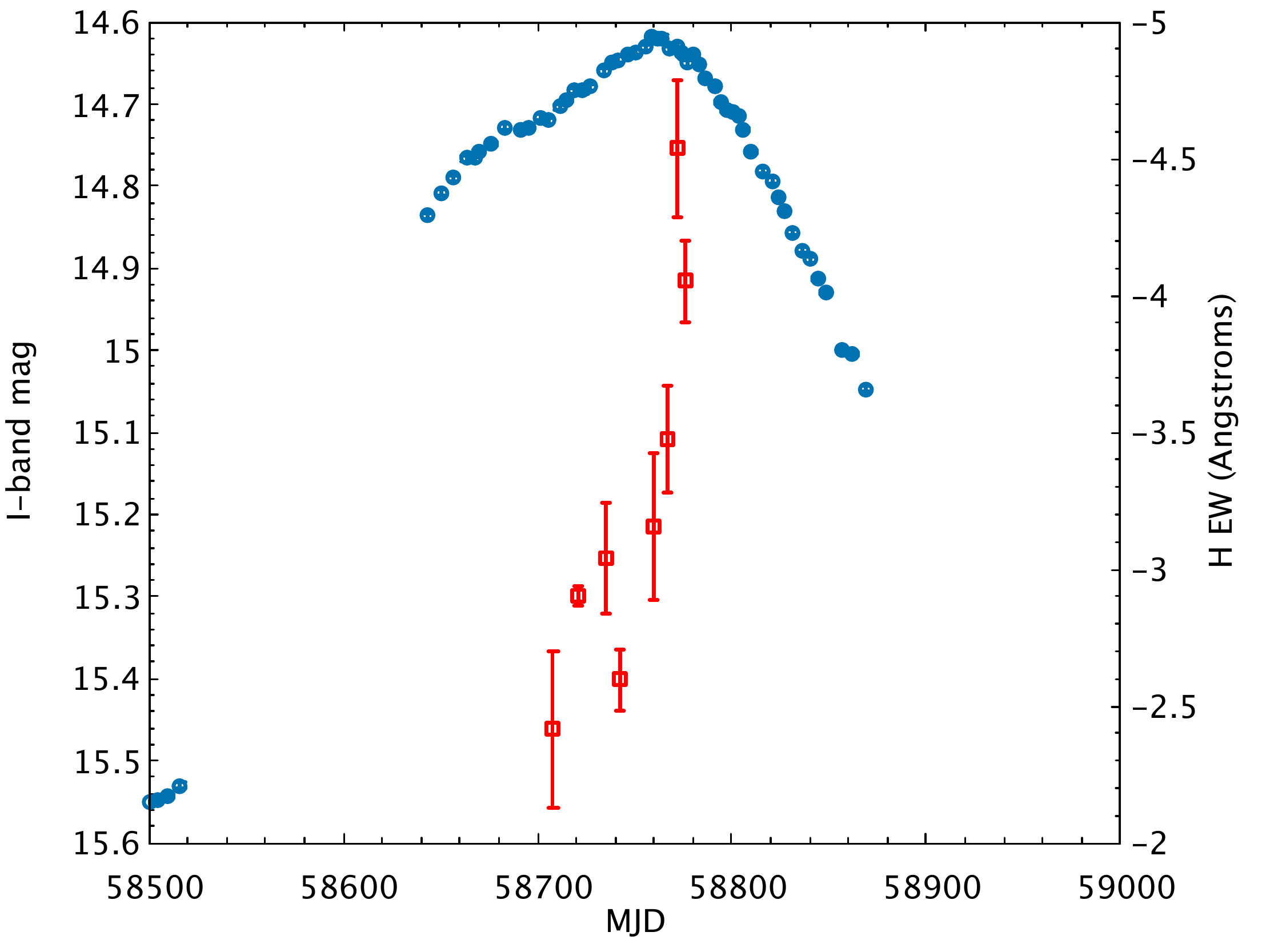}}
\caption{Evolution of the H$\alpha$ EW during the current OGLE outburst. The blue circles represent the OGLE data and the red squares represent the H$\alpha$ EW measurments.}
\label{fig:EW_evolution}
\end{figure}

\begin{figure}
\resizebox{\hsize}{!}
            {\includegraphics[angle=0,width=20cm]{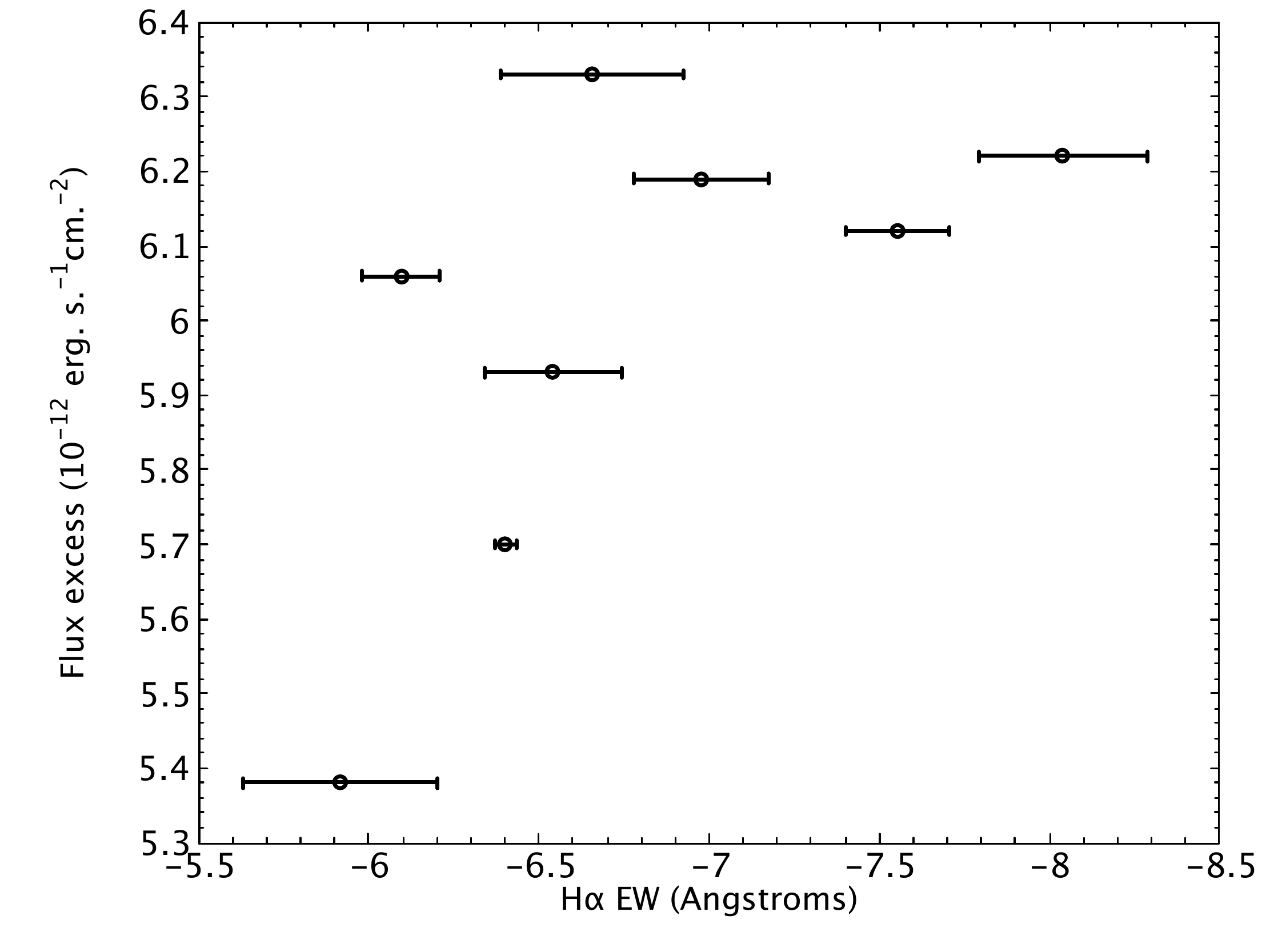}}
\caption{Correlation between the total H$\alpha$ EW and excess $I$-band flux for measurements taken on the same date. As discussed in the text the H$\alpha$ values shown here include an extra component needed to fill in the photospheric line and the I-band flux is the excess over the disc-less state.}
\label{fig:EW_flux_correlation}
\end{figure}

In Fig.~\ref{fig:EW_evolution} we show the evolution of the H$\alpha$ EW during the current $I-$band outburst, which, in Be stars, is used as an indicator of the size of the disc. Fig.~\ref{fig:EW_flux_correlation} shows the correlation between the H$\alpha$ EW and the $I$-band excess flux for measurements obtained on the same date to demonstrate the spectroscopic and photometric relationship on a linear scale. The excess flux is the flux emitted by the disc and is obtained by subtracting the stellar $I$-band flux (converted from the base magnitude of $I \sim 15.6$ mag.) from the OGLE $I$-band flux.  We added a value of -3.5 \AA\ to our measured values to account for the infilling of the absorption line by the Be disc during its formation (the H$\alpha$ EW value of the absorption line for a B0 star has a value 3.5 \AA; \citealt{1987clst.book.....J}). The EW measurements show an increasing trend throughout the course of the monitoring campaign.  The EW values are relatively low, given that the SALT observations were obtained close to the peak of the OGLE outburst. In a simple picture, one might expect the unusually large amplitude of the OGLE lightcurve to correspond to a very large disc, the emission of which would be evident in the EW measurements. We discuss this further in Sect.~\ref{sec:discussion}.\\

\begin{table}
	\centering
	\caption{A log of all the SALT observations. The H$\alpha$ equivalent width measurements of X0103 are presented in this table. The spectrum taken with the PG1300 grating only covers the blue region, hence no H$\alpha$ equivalent width measurement is recorded for that observation.}
	\label{tab:EW_measurements}
    \setlength\tabcolsep{2pt}
	\begin{tabular}{ccccc} 
		\hline\hline
		Date & MJD & EW (\AA) & Grating &  S/N  \\
		\hline

12 August 2019 & 58708.02 &	-2.42 $\pm$ 0.29 & PG0900 &  105 \\
25 August 2019 & 58721.08 &	-2.902 $\pm$ 0.034 & PG1800 &  135 \\
8 September 2019 & 58734.99  & $-$ & PG1300 &  108 \\
8 September 2019 & 58735.01 &	-3.04 $\pm$ 0.20 & PG1800 &  65 \\
15 September 2019 &58742.06 &	-2.60 $\pm$ 0.11 & PG1800 &  88 \\
03 October 2019 & 58759.89 &	-3.16 $\pm$ 0.27 & PG1800 &  105 \\
10 October 2019 & 58766.88 &	-3.47 $\pm$ 0.20 & PG1800 &  85 \\
15 October 2019 & 58771.85 &	-4.54 $\pm$ 0.25 & PG1800 &  95 \\
19 October 2019 & 58775.87 &	-4.05 $\pm$ 0.15 & PG2300 &  90 \\

		\hline
	\end{tabular}
\label{tab:SALT}	
\end{table}

\subsection{X-ray observations}
 The first X-ray detection from S-CUBED is around the time of the start of the optical rise, accompanied by a few more during the rise of the optical outburst, with three more detections seen at the end of the outburst when the optical flux was at a minimum. In the current optical outburst no X-ray detections are seen at the start of the outburst, in contrast to the previous one, but a few more are seen during its rise. The X-ray flux from these detections is relatively low, with a range $(2-14) \times 10^{-11}$~erg.s$^{-1}$cm$^{-2}$ ($L_X \sim (9.2-55)\times 10^{35}$~erg.s$^{-1}$). The flux from a longer \textit{Swift} observation ($\sim 2$~ks) during the rise of the current optical outburst measures at a lower value of $\sim 8.7\times 10^{-14}$~erg.s$^{-1}$cm$^{-2}$ ($L_X \sim 4\times 10^{34}$~erg.s$^{-1}$). The source has been observed previously in the X-rays using \textit{XMM-}Newton during optical minimum and rise with flux measurements showing similar values on the order of $10^{-14}$~erg.s$^{-1}$cm$^{-2}$ \citep{2013A&A...558A...3S}. One would expect, naively, the X-ray emission to be correlated with the photometric flux as is typically observed in other BeXBs (e.g. \citealt{2011MNRAS.413.1600R, 2019MNRAS.489..993M}), hence it is surprising the low X-ray flux observed during the recent optical outburst.

\section{Discussion}
\label{sec:discussion}
\subsection{Low H$\alpha$ emission from the Be disc during a large optical outburst}
In this sub-section we discuss the possible explanations for the surprisingly low H$\alpha$ emission during the peak of the photometric outburst in the context of disc truncation and density variability of the Be disc.

\subsubsection{Disc size changes and truncation}
The peak separation of double-peak profiles of the H$\alpha$ emission line originating from Be discs has been shown to give an idea of the physical extent of the disc \citep{1972ApJ...171..549H}. We did not use this method of calculating the disc size, as some of the H$\alpha$ line profiles display single peaks . \cite{1994A&A...289..458H} demonstrates that using the peak separation of the H$\alpha$ line is a more reliable estimate in the case of optically thin discs. The EW of the H$\alpha$ emission line is also used as a gauge to the size of the disc and was shown to be correlated to radii measurements from optical interferometry of nearby isolated Be stars \citep{2006ApJ...651L..53G}. In BeXBs \cite{1997A&A...322..193R} show that the maximum EW ever recorded for different systems correlates with the orbital period, which is interpreted as an indicator for disc truncation. Disc truncation in BeXBs can be explained as a consequence of the viscous decretion disc model, where the NS acts as a barrier that limits the extent to which the disc can grow \citep{1991MNRAS.250..432L,2001A&A...377..161O}. The truncation of the disc by the NS results in Be discs that are thicker and of enhanced surface density in BeXBs compared to isolated Be stars, with the density dropping rapidly as one moves outwards \citep{2002MNRAS.337..967O}. \\


The amplitudes of the optical outbursts in X0103 (top and middle panels of Fig.~\ref{fig:ogle}) are the largest ever observed in a BeXB (see \citealt{2011MNRAS.413.1600R}, for example, for a study of the largest sample of BeXBs using photometric datasets). Assuming that the increase in continuum emission originates from the disc, then one would expect to find larger H$\alpha$ EW measurements at OGLE maxima than you would at OGLE minima since the EW has a faster rate of change close to OGLE maxima (see section~\ref{sec:disc_opacity}) .\\

\subsubsection{Disc opacity and density changes}
\label{sec:disc_opacity}
A possible explanation for the low H$\alpha$ EW measurements during the large optical outburst could be that the disc has a rather steep density distribution where continuum flux originates from the inner regions of the disc that are optically thick and do not contribute much line emission. As one moves towards the outer regions of the disc, where it becomes optically thinner, the line emission is more dominant. In such a scenario, it is possible for the radial extent of the disc to be relatively small, as seen in the low EW measurements, while the continuum flux is high.   
The colour-magnitude plots in Fig.~\ref{fig:col_mag} show the flux peaking at $(V-I) \approx 0$, i.e. when the blue and red emission contribution is equal. The range in colour from when the mass outflow from the Be star starts to when it stops at maximum flux (phase "I" in Fig.~\ref{fig:col_mag}) is larger than when mass outflow stops to when the optically thick region is completely removed (phase "II"). This change in flux with colour shows that the majority of the continuum emission originates from the inner regions of the disc which are optically thick. 

From Fig.~\ref{fig:col_mag} (a) phase "II" shows a steeper slope (i.e. a more rapid decline in continuum flux over the same colour range) compared to that of Fig.~\ref{fig:col_mag} (b). This rapid decrease in $V$-band flux indicates that the optically thick blue region (inner disc) is removed faster than the red region once the mass outflow from the Be star stops. The slopes during phase "III" (i.e. when the flux is decreasing while the colour is becoming blue) show a steeper change in Fig.~\ref{fig:col_mag} (b) than Fig.~\ref{fig:col_mag} (a) since this is the period when the optically thin material from the disc is being removed, where the red emission dominates. \\
The H$\alpha$ EW measurements were obtained close to the peak of the optical outburst and are seen to change at a faster rate than the $I$-band flux during this time (a change in the EW of $\sim$35\% compared to $\sim$15\% of the $I$-band flux; Fig.~\ref{fig:EW_flux_correlation}). A possible reason for the faster change in EW compared to the $I$-band flux during this time could be that close to the peak only the outer and more optically thin parts of the disc, which contribute more to the line emission and less to the continuum, are growing more rapidly.\\
The low X-ray emission seen in the \textit{Swift} observations support this idea of a radially small disc which does not extend to a large-enough region for the NS to accrete a significant amount of matter. \\

\subsection{Quasi-periodicity of the optical outbursts}
In this sub-section we discuss different possibilities for the origin of the periodicity seen in the optical outbursts from OGLE data.

\subsubsection{Is the recurrence time of the optical outbursts due to the orbital period?}
Orbital periods on the order of the size of the peak separation of the optical outbursts seen in X0103 ($\sim$1200 days; Fig.~\ref{fig:ogle}) are generally rare in BeXBs. \cite{2019ApJ...884....2L} suggested that the $\sim$1180 day peak separation of the OGLE outbursts in the BeXB system, CXO J005215.4-731915, to be attributable to the orbital period. This is in contrast to earlier analysis performed by \cite{2011MNRAS.412..391S} on the same source, where it was suggested that the long-term periodicity is due to structural changes in the disc. Similarly, the SMC BeXB candidate, Swift J010745.0-722740, was seen to display optical outbursts separated by $\sim$1180 days which are accompanied by strong X-ray emission around the optical peaks, suggesting an orbital period of the system \citep{2014ATel.5778....1M,2017ATel10253....1V}. The similar $\sim$1200 day optical peak separation in X0103 may therefore be suggestive of an orbital period. The X-ray emission from the \textit{Swift} observations, however, makes this claim difficult to reconcile with, as the emission is relatively weak throughout its monitoring, being just at the detectable limit even close to the optical peak.\\
Moreover, the maximum H$\alpha$ EW measured in this work suggests that the orbital period is short (on the order of $\sim$30~days) when interpolated on the correlation plot between the maximum historical H$\alpha$ EW and orbital period in BeXBs \citep{1997A&A...322..193R}.  

\subsubsection{Periodic mass-loss from the Be star}
BeXBs are also known to display superorbital periods on long timescales ($\sim$200-3000 days) and this long-term variability is attributed to the growth and depletion of the disc \citep{2011MNRAS.413.1600R}. A possible mechanism for the long-term variability seen in the optical variability of X0103 could be a combination of the mass-loss episodes of the Be star giving rise to the disc and a co-planar geometry of the NS orbit relative to the disc plane. In this scenario, the NS orbits the Be star/disc in a roughly circular orbit, resulting in efficient truncation of the disc \citep{2001A&A...377..161O}. The NS then acts as an impediment, limiting the disc growth, resulting in the inner regions to be vertically thick and dense due to a large accumulation of matter. The mass-loss from the Be star switching off then results in the dissipation of the disc (and hence a drop in flux), where the inner regions are lost first. This is what ultimately gives rise to the loop structure in the colour-magnitude plot in Fig.~\ref{fig:col_mag}: the switching off of mass-loss is required at peak emission in Fig.~\ref{fig:col_mag} for phase II to occur (i.e. the period when the inner, optically thick, blue regions dissipate faster than the redder, outer regions).\\

\section{Conclusions}
\label{sec:conclusion}
We have presented recent and long-term variability of the Be X-ray binary system, X0103. The long-term optical photometric data shows unusually large outbursts due to Be disc variability. We confirm the BeXB nature of the system though the presence of the H$\alpha$ line in emission. The blue spectrum from SALT allows us to classify the spectral type of the optical companion as O9.7 - B0 IV-Ve. The H$\alpha$ EW and X-ray flux display relatively low emission, considering that the amplitude of the photometric outubrsts is one of the largest from the known BeXB systems. Using the optical colour variability, we explain this unusual behaviour though the changes in the density and opacity of the disc rather than the radial extent. We consider the origin of the long quasi-periodicity of the optical outbursts, which we propose is due to the geometry of the system and periodic mass-loss events from the Be star.

\section*{Acknowledgements}

IMM, DAHB and VAM are supported by the South African NRF. Some of the observations reported in this paper were obtained with the Southern African Large Telescope (SALT), as part of the Large Science Programme on transients 2018-2-LSP-001 (PI: Buckley). The OGLE project has received funding from the National
Science Centre, Poland, grant MAESTRO 2014/14/A/ST9/00121
to AU. IN is partially supported by the Spanish Government under grant PGC2018-093741-B-C21 (MICIU/AEI/FEDER, UE).
Polish participation in SALT is funded by grant No. MNiSW DIR/WK/2016/07.




\bibliographystyle{mnras}
\bibliography{ref} 

\begin{thebibliography}{}
\makeatletter
\relax
\def\mn@urlcharsother{\let\do\@makeother \do\$\do\&\do\#\do\^\do\_\do\%\do\~}
\def\mn@doi{\begingroup\mn@urlcharsother \@ifnextchar [ {\mn@doi@}
  {\mn@doi@[]}}
\def\mn@doi@[#1]#2{\def\@tempa{#1}\ifx\@tempa\@empty \href
  {http://dx.doi.org/#2} {doi:#2}\else \href {http://dx.doi.org/#2} {#1}\fi
  \endgroup}
\def\mn@eprint#1#2{\mn@eprint@#1:#2::\@nil}
\def\mn@eprint@arXiv#1{\href {http://arxiv.org/abs/#1} {{\tt arXiv:#1}}}
\def\mn@eprint@dblp#1{\href {http://dblp.uni-trier.de/rec/bibtex/#1.xml}
  {dblp:#1}}
\def\mn@eprint@#1:#2:#3:#4\@nil{\def\@tempa {#1}\def\@tempb {#2}\def\@tempc
  {#3}\ifx \@tempc \@empty \let \@tempc \@tempb \let \@tempb \@tempa \fi \ifx
  \@tempb \@empty \def\@tempb {arXiv}\fi \@ifundefined
  {mn@eprint@\@tempb}{\@tempb:\@tempc}{\expandafter \expandafter \csname
  mn@eprint@\@tempb\endcsname \expandafter{\@tempc}}}

\bibitem[\protect\citeauthoryear{{Buckley}, {Swart}  \& {Meiring}}{{Buckley}
  et~al.}{2006}]{2006SPIE.6267E..0ZB}
{Buckley} D.~A.~H.,  {Swart} G.~P.,   {Meiring} J.~G.,  2006, in Society of
  Photo-Optical Instrumentation Engineers (SPIE) Conference Series. p. 62670Z,
  \mn@doi{10.1117/12.673750}

\bibitem[\protect\citeauthoryear{{Burgh}, {Nordsieck}, {Kobulnicky},
  {Williams}, {O'Donoghue}, {Smith}  \& {Percival}}{{Burgh}
  et~al.}{2003}]{2003SPIE.4841.1463B}
{Burgh} E.~B.,  {Nordsieck} K.~H.,  {Kobulnicky} H.~A.,  {Williams} T.~B.,
  {O'Donoghue} D.,  {Smith} M.~P.,   {Percival} J.~W.,  2003, in {Iye} M.,
  {Moorwood} A.~F.~M.,  eds,  \procspie Vol. 4841, Instrument Design and
  Performance for Optical/Infrared Ground-based Telescopes. pp 1463--1471,
  \mn@doi{10.1117/12.460312}

\bibitem[\protect\citeauthoryear{{Coleiro} \& {Chaty}}{{Coleiro} \&
  {Chaty}}{2013}]{2013ApJ...764..185C}
{Coleiro} A.,  {Chaty} S.,  2013, \mn@doi [\apj] {10.1088/0004-637X/764/2/185},
  \href {https://ui.adsabs.harvard.edu/abs/2013ApJ...764..185C} {764, 185}

\bibitem[\protect\citeauthoryear{{Collins}}{{Collins}}{1987}]{1987pbes.coll....3C}
{Collins} George~W. I.,  1987, in {Slettebak} A.,  {Snow} T.~P.,  eds, IAU
  Colloq. 92: Physics of Be Stars. p.~3

\bibitem[\protect\citeauthoryear{{Cranmer}}{{Cranmer}}{2009}]{2009ApJ...701..396C}
{Cranmer} S.~R.,  2009, \mn@doi [\apj] {10.1088/0004-637X/701/1/396}, \href
  {https://ui.adsabs.harvard.edu/abs/2009ApJ...701..396C} {701, 396}

\bibitem[\protect\citeauthoryear{{Crawford} et~al.,}{{Crawford}
  et~al.}{2012}]{2012ascl.soft07010C}
{Crawford} S.~M.,  et~al., 2012, {PySALT: SALT science pipeline}, Astrophysics
  Source Code Library (\mn@eprint {ascl} {1207.010})

\bibitem[\protect\citeauthoryear{{Dickey} \& {Lockman}}{{Dickey} \&
  {Lockman}}{1990}]{1990ARA&A..28..215D}
{Dickey} J.~M.,  {Lockman} F.~J.,  1990, \mn@doi [\araa]
  {10.1146/annurev.aa.28.090190.001243}, \href
  {https://ui.adsabs.harvard.edu/abs/1990ARA&A..28..215D} {28, 215}

\bibitem[\protect\citeauthoryear{{Evans} et~al.,}{{Evans}
  et~al.}{2009}]{2009MNRAS.397.1177E}
{Evans} P.~A.,  et~al., 2009, \mn@doi [\mnras]
  {10.1111/j.1365-2966.2009.14913.x}, \href
  {https://ui.adsabs.harvard.edu/abs/2009MNRAS.397.1177E} {397, 1177}

\bibitem[\protect\citeauthoryear{{Filippova}, {Tsygankov}, {Lutovinov}  \&
  {Sunyaev}}{{Filippova} et~al.}{2005}]{2005AstL...31..729F}
{Filippova} E.~V.,  {Tsygankov} S.~S.,  {Lutovinov} A.~A.,   {Sunyaev} R.~A.,
  2005, \mn@doi [Astronomy Letters] {10.1134/1.2123288}, \href
  {https://ui.adsabs.harvard.edu/abs/2005AstL...31..729F} {31, 729}

\bibitem[\protect\citeauthoryear{{Gehrels} et~al.,}{{Gehrels}
  et~al.}{2004}]{2004ApJ...611.1005G}
{Gehrels} N.,  et~al., 2004, \mn@doi [\apj] {10.1086/422091}, \href
  {http://adsabs.harvard.edu/abs/2004ApJ...611.1005G} {611, 1005}

\bibitem[\protect\citeauthoryear{{Graczyk}, {Pietrzy{\'n}ski}, {Pilecki},
  {Thompson}, {Gieren}, {Konorski}, {Udalski}  \& {Soszy{\'n}ski}}{{Graczyk}
  et~al.}{2013}]{2013IAUS..289..222G}
{Graczyk} D.,  {Pietrzy{\'n}ski} G.,  {Pilecki} B.,  {Thompson} I.~B.,
  {Gieren} W.,  {Konorski} P.,  {Udalski} A.,   {Soszy{\'n}ski} I.,  2013, in
  {de Grijs} R.,  ed.,  IAU Symposium Vol. 289, Advancing the Physics of Cosmic
  Distances. pp 222--225 (\mn@eprint {arXiv} {1311.1270}),
  \mn@doi{10.1017/S1743921312021436}

\bibitem[\protect\citeauthoryear{{Grundstrom} \& {Gies}}{{Grundstrom} \&
  {Gies}}{2006}]{2006ApJ...651L..53G}
{Grundstrom} E.~D.,  {Gies} D.~R.,  2006, \mn@doi [\apjl] {10.1086/509635},
  \href {https://ui.adsabs.harvard.edu/abs/2006ApJ...651L..53G} {651, L53}

\bibitem[\protect\citeauthoryear{{Haberl} \& {Sturm}}{{Haberl} \&
  {Sturm}}{2016}]{2016A&A...586A..81H}
{Haberl} F.,  {Sturm} R.,  2016, \mn@doi [\aap] {10.1051/0004-6361/201527326},
  \href {https://ui.adsabs.harvard.edu/abs/2016A&A...586A..81H} {586, A81}

\bibitem[\protect\citeauthoryear{{Harmanec}}{{Harmanec}}{1983}]{1983HvaOB...7...55H}
{Harmanec} P.,  1983, Hvar Observatory Bulletin, \href
  {http://adsabs.harvard.edu/abs/1983HvaOB...7...55H} {7, 55}

\bibitem[\protect\citeauthoryear{{Huang}}{{Huang}}{1972}]{1972ApJ...171..549H}
{Huang} S.-S.,  1972, \mn@doi [\apj] {10.1086/151309}, \href
  {https://ui.adsabs.harvard.edu/abs/1972ApJ...171..549H} {171, 549}

\bibitem[\protect\citeauthoryear{{Hummel}}{{Hummel}}{1994}]{1994A&A...289..458H}
{Hummel} W.,  1994, \aap, \href
  {https://ui.adsabs.harvard.edu/abs/1994A&A...289..458H} {289, 458}

\bibitem[\protect\citeauthoryear{{Jaschek} \& {Jaschek}}{{Jaschek} \&
  {Jaschek}}{1987}]{1987clst.book.....J}
{Jaschek} C.,  {Jaschek} M.,  1987, {The classification of stars}

\bibitem[\protect\citeauthoryear{{Kennea}, {Coe}, {Evans}, {Waters}  \&
  {Jasko}}{{Kennea} et~al.}{2018}]{2018ApJ...868...47K}
{Kennea} J.~A.,  {Coe} M.~J.,  {Evans} P.~A.,  {Waters} J.,   {Jasko} R.~E.,
  2018, \mn@doi [\apj] {10.3847/1538-4357/aae839}, \href
  {http://adsabs.harvard.edu/abs/2018ApJ...868...47K} {868, 47}

\bibitem[\protect\citeauthoryear{{Kobulnicky}, {Nordsieck}, {Burgh}, {Smith},
  {Percival}, {Williams}  \& {O'Donoghue}}{{Kobulnicky}
  et~al.}{2003}]{2003SPIE.4841.1634K}
{Kobulnicky} H.~A.,  {Nordsieck} K.~H.,  {Burgh} E.~B.,  {Smith} M.~P.,
  {Percival} J.~W.,  {Williams} T.~B.,   {O'Donoghue} D.,  2003, in {Iye} M.,
  {Moorwood} A.~F.~M.,  eds,  \procspie Vol. 4841, Instrument Design and
  Performance for Optical/Infrared Ground-based Telescopes. pp 1634--1644,
  \mn@doi{10.1117/12.460315}

\bibitem[\protect\citeauthoryear{{Lazzarini} et~al.,}{{Lazzarini}
  et~al.}{2019}]{2019ApJ...884....2L}
{Lazzarini} M.,  et~al., 2019, \mn@doi [\apj] {10.3847/1538-4357/ab3f32}, \href
  {https://ui.adsabs.harvard.edu/abs/2019ApJ...884....2L} {884, 2}

\bibitem[\protect\citeauthoryear{{Lee}, {Osaki}  \& {Saio}}{{Lee}
  et~al.}{1991}]{1991MNRAS.250..432L}
{Lee} U.,  {Osaki} Y.,   {Saio} H.,  1991, \mn@doi [\mnras]
  {10.1093/mnras/250.2.432}, \href
  {https://ui.adsabs.harvard.edu/abs/1991MNRAS.250..432L} {250, 432}

\bibitem[\protect\citeauthoryear{{Maggi}, {Sturm}, {Haberl}, {Vasilopoulos}  \&
  {Udalski}}{{Maggi} et~al.}{2014}]{2014ATel.5778....1M}
{Maggi} P.,  {Sturm} R.,  {Haberl} F.,  {Vasilopoulos} G.,   {Udalski} A.,
  2014, The Astronomer's Telegram, \href
  {https://ui.adsabs.harvard.edu/abs/2014ATel.5778....1M} {5778, 1}

\bibitem[\protect\citeauthoryear{{Monageng} et~al.,}{{Monageng}
  et~al.}{2019}]{2019MNRAS.489..993M}
{Monageng} I.~M.,  et~al., 2019, \mn@doi [\mnras] {10.1093/mnras/stz2262},
  \href {https://ui.adsabs.harvard.edu/abs/2019MNRAS.489..993M} {489, 993}

\bibitem[\protect\citeauthoryear{{Okazaki} \& {Negueruela}}{{Okazaki} \&
  {Negueruela}}{2001}]{2001A&A...377..161O}
{Okazaki} A.~T.,  {Negueruela} I.,  2001, \mn@doi [\aap]
  {10.1051/0004-6361:20011083}, \href
  {https://ui.adsabs.harvard.edu/abs/2001A&A...377..161O} {377, 161}

\bibitem[\protect\citeauthoryear{{Okazaki}, {Bate}, {Ogilvie}  \&
  {Pringle}}{{Okazaki} et~al.}{2002}]{2002MNRAS.337..967O}
{Okazaki} A.~T.,  {Bate} M.~R.,  {Ogilvie} G.~I.,   {Pringle} J.~E.,  2002,
  \mn@doi [\mnras] {10.1046/j.1365-8711.2002.05960.x}, \href
  {https://ui.adsabs.harvard.edu/abs/2002MNRAS.337..967O} {337, 967}

\bibitem[\protect\citeauthoryear{{Porter} \& {Rivinius}}{{Porter} \&
  {Rivinius}}{2003}]{2003PASP..115.1153P}
{Porter} J.~M.,  {Rivinius} T.,  2003, \mn@doi [\pasp] {10.1086/378307}, \href
  {https://ui.adsabs.harvard.edu/abs/2003PASP..115.1153P} {115, 1153}

\bibitem[\protect\citeauthoryear{{Rajoelimanana}, {Charles}  \&
  {Udalski}}{{Rajoelimanana} et~al.}{2011}]{2011MNRAS.413.1600R}
{Rajoelimanana} A.~F.,  {Charles} P.~A.,   {Udalski} A.,  2011, \mn@doi
  [\mnras] {10.1111/j.1365-2966.2011.18243.x}, \href
  {http://adsabs.harvard.edu/abs/2011MNRAS.413.1600R} {413, 1600}

\bibitem[\protect\citeauthoryear{{Reig}}{{Reig}}{2011}]{2011Ap&SS.332....1R}
{Reig} P.,  2011, \mn@doi [\apss] {10.1007/s10509-010-0575-8}, \href
  {https://ui.adsabs.harvard.edu/abs/2011Ap&SS.332....1R} {332, 1}

\bibitem[\protect\citeauthoryear{{Reig} \& {Fabregat}}{{Reig} \&
  {Fabregat}}{2015}]{2015A&A...574A..33R}
{Reig} P.,  {Fabregat} J.,  2015, \mn@doi [\aap] {10.1051/0004-6361/201425008},
  \href {http://adsabs.harvard.edu/abs/2015A%26A...574A..33R} {574, A33}

\bibitem[\protect\citeauthoryear{{Reig}, {Fabregat}  \& {Coe}}{{Reig}
  et~al.}{1997}]{1997A&A...322..193R}
{Reig} P.,  {Fabregat} J.,   {Coe} M.~J.,  1997, \aap, \href
  {https://ui.adsabs.harvard.edu/abs/1997A&A...322..193R} {322, 193}

\bibitem[\protect\citeauthoryear{{Schurch}, {Coe}, {McBride}, {Townsend},
  {Udalski}, {Haberl}  \& {Corbet}}{{Schurch}
  et~al.}{2011}]{2011MNRAS.412..391S}
{Schurch} M.~P.~E.,  {Coe} M.~J.,  {McBride} V.~A.,  {Townsend} L.~J.,
  {Udalski} A.,  {Haberl} F.,   {Corbet} R.~H.~D.,  2011, \mn@doi [\mnras]
  {10.1111/j.1365-2966.2010.17914.x}, \href
  {https://ui.adsabs.harvard.edu/abs/2011MNRAS.412..391S} {412, 391}

\bibitem[\protect\citeauthoryear{{Stella}, {White}  \& {Rosner}}{{Stella}
  et~al.}{1986}]{1986ApJ...308..669S}
{Stella} L.,  {White} N.~E.,   {Rosner} R.,  1986, \mn@doi [\apj]
  {10.1086/164538}, \href {http://adsabs.harvard.edu/abs/1986ApJ...308..669S}
  {308, 669}

\bibitem[\protect\citeauthoryear{{Straizys} \& {Kuriliene}}{{Straizys} \&
  {Kuriliene}}{1981}]{1981Ap&SS..80..353S}
{Straizys} V.,  {Kuriliene} G.,  1981, \mn@doi [\apss] {10.1007/BF00652936},
  \href {https://ui.adsabs.harvard.edu/abs/1981Ap&SS..80..353S} {80, 353}

\bibitem[\protect\citeauthoryear{{Sturm} et~al.,}{{Sturm}
  et~al.}{2013}]{2013A&A...558A...3S}
{Sturm} R.,  et~al., 2013, \mn@doi [\aap] {10.1051/0004-6361/201219935}, \href
  {https://ui.adsabs.harvard.edu/abs/2013A&A...558A...3S} {558, A3}

\bibitem[\protect\citeauthoryear{{Udalski}, {Kubiak}  \& {Szymanski}}{{Udalski}
  et~al.}{1997}]{1997AcA....47..319U}
{Udalski} A.,  {Kubiak} M.,   {Szymanski} M.,  1997, \actaa, \href
  {http://ukads.nottingham.ac.uk/abs/1997AcA....47..319U} {47, 319}

\bibitem[\protect\citeauthoryear{{Udalski}, {Szyma{\'n}ski}  \&
  {Szyma{\'n}ski}}{{Udalski} et~al.}{2015}]{2015AcA....65....1U}
{Udalski} A.,  {Szyma{\'n}ski} M.~K.,   {Szyma{\'n}ski} G.,  2015, \actaa,
  \href {http://ukads.nottingham.ac.uk/abs/2015AcA....65....1U} {65, 1}

\bibitem[\protect\citeauthoryear{{Vasilopoulos}, {Haberl}  \&
  {Maggi}}{{Vasilopoulos} et~al.}{2017}]{2017ATel10253....1V}
{Vasilopoulos} G.,  {Haberl} F.,   {Maggi} P.,  2017, The Astronomer's
  Telegram, \href {https://ui.adsabs.harvard.edu/abs/2017ATel10253....1V}
  {10253, 1}

\bibitem[\protect\citeauthoryear{{Wegner}}{{Wegner}}{1993}]{1993AcA....43..209W}
{Wegner} W.,  1993, \actaa, \href
  {https://ui.adsabs.harvard.edu/abs/1993AcA....43..209W} {43, 209}

\bibitem[\protect\citeauthoryear{{de Wit}, {Lamers}, {Marquette}  \&
  {Beaulieu}}{{de Wit} et~al.}{2006}]{2006A&A...456.1027D}
{de Wit} W.~J.,  {Lamers} H.~J.~G.~L.~M.,  {Marquette} J.~B.,   {Beaulieu}
  J.~P.,  2006, \mn@doi [\aap] {10.1051/0004-6361:20065137}, \href
  {https://ui.adsabs.harvard.edu/abs/2006A&A...456.1027D} {456, 1027}

\makeatother
\end{thebibliography}


\begin{thebibliography}{99}
\bibitem[\protect\citeauthoryear{Author}{2012}]{Author2012}
Author A.~N., 2013, Journal of Improbable Astronomy, 1, 1
\bibitem[\protect\citeauthoryear{Others}{2013}]{Others2013}
Others S., 2012, Journal of Interesting Stuff, 17, 198
\end{thebibliography}

\bsp	
\label{lastpage}
\end{document}